\documentclass[aps,prl,preprint,groupedaddress]{revtex4-2}

\usepackage{amssymb}
\usepackage{graphicx}
\usepackage{amsmath}
\usepackage{xcolor}

\begin{document}


\title{Photonic Shankar skyrmion}


\author{Haiwen Wang}
\email[]{hwwang@stanford.edu}
\affiliation{Department of Applied Physics, Stanford University, Stanford, CA, 94305, USA}

\author{Shanhui Fan}
\email[]{shanhui@stanford.edu}
\affiliation{Department of Electrical Engineering, Stanford University, Stanford, CA, 94305, USA}


\date{\today}

\begin{abstract}
We unveil a new topological quasiparticle of light in 3D space, named the photonic Shankar skyrmion. We show that an elliptically polarized field can be described by an SO(3) order parameter and it can form a texture in 3D space classified by the $\pi_3(SO(3))$ homotopy group, known as the Shankar skyrmion. We provide ways to construct the photonic Shankar skyrmion in static monochromatic waves and also in propagating wavepackets, the latter give rise to a flying topological quasiparticle. We demonstrate that the transition between the topological configuration and the trivial configuration gives rise to a novel topological singularity, which we call the $L^T$ surface. Such configurations of the electromagnetic field, under light-matter interaction, may lead to new phenomena in condensed matter physics and plasma physics, and are expected to find applications in quantum emulation and optical manipulation.
\end{abstract}


\maketitle


Topological field configurations, i.e., field configurations with non-trivial topologies, have been of fundamental interest in various areas of physics. These configurations have been used for explaining the structures of fundamental particles. Examples include Lord Kelvin's vortex atom hypothesis \cite{kelvin1867vortex}, the skyrmion as proposed by Skyrme \cite{skyrme1961non, skyrme1962unified}, and more recent works on topological solitons \cite{faddeev1997stable, sutcliffe2007knots, manton2004topological}. The topological field configurations can be mathematically classified by homotopy groups \cite{mermin1979topological}, which assign discrete topological charges to the field configurations. Under small perturbations, the topological charges will not change, and therefore, such field configurations exhibit robustness.

Many topological field configurations have been found in various condensed matter systems. For example, the baby skyrmions, or two-dimensional (2D) skyrmions, which refer to the localized distributions on a 2D plane of an order parameter consisting of unit 3-vectors, have been observed in magnetic materials \cite{yu2010real, muhlbauer2009skyrmion, han2017skyrmions} and liquid crystals \cite{foster2019two}. 2D skyrmions are classified by the second homotopy group $\pi_2(S^2)$. Here $S^2$ refers to the 2-sphere where the order parameter takes value. In three-dimensional (3D) space, a variety of topological field configurations are possible. These include hopfions in magnetic materials \cite{kent2021creation, zheng2023hopfion} and liquid crystals \cite{ackerman2015self, tai2019three, wu2022hopfions}, skyrmions in two-component Bose-Einstein Condensates (BECs) \cite{kawakami2012stable}, and Shankar skyrmions in superfluid $^3$He-A \cite{shankar1977applications, volovik1977particle} and ferromagnetic spin-1 BECs \cite{al2001skyrmions, lee2018synthetic}. These field configurations are classified by the third homotopy groups $\pi_3(S^2)$, $\pi_3(S^3)$, and $\pi_3(SO(3))$ respectively. Mechanical waves in water have also been shown to form topological field configurations \cite{smirnova2024water}. Topological configurations of fields generally vary in a localized region of space, and have been shown to display particle-like dynamics \cite{wang2019current, poy2022interaction} or to form macromolecules and crystals \cite{ackerman2015self, tai2019three, zhao2023topological}. Thus, such configurations are often viewed as topological quasiparticles.

In recent years, there have been substantial interests in creating topological field configurations using electromagnetic waves. These configurations are sometimes referred to as the topological quasiparticles of light \cite{shen2024optical}. Topological configurations of baby skyrmions \cite{tsesses2018optical, du2019deep, davis2020ultrafast, shen2021supertoroidal, lei2021photonic, ornelas2024non}, hopfions \cite{wang2023photonic, shen2023topological, wu2024photonic}, and skyrmions \cite{sugic2021particle, ehrmanntraut2023optical} have been demonstrated with electromagnetic waves. These configurations can exist in free space \cite{shen2025free} and propagate at the speed of light \cite{wan2022scalar, lyu2024formation}. Such topological quasiparticles of light may lead to novel excitations of matter \cite{parmee2022optical, gao2024dynamical} or enable new applications in light-matter interactions \cite{yang2023spin}.

In this Letter, we demonstrate the capability of electromagnetic waves to form Shankar skyrmions, a possibility that has never been discussed before. We find that an SO(3) order parameter can be defined using elliptically polarized electric fields, achieving the same order parameter as superfluid $^3$He-A \cite{shankar1977applications, volovik1977particle} and ferromagnetic spin-1 Bose-Einstein condensate (BEC) \cite{al2001skyrmions, lee2018synthetic}. Since the homotopy group $\pi_3(SO(3))=\mathbb{Z}$, a topological field configuration for an SO(3) order parameter can potentially exist in 3D space, and we verify that it can be created as a solution of Maxwell's equations in free space. We further investigate the topological defect associated with such texture, and we find that the defects appear as a 2D surface, which we name the $L^T$ surface. Such topological quasiparticles of light can propagate in free space. We show that by engineering the spatiotemporal correlation of the electromagnetic waves, the velocity of such topological quasiparticles can be arbitrarily controlled. Photonic Shankar skyrmions may lead to interesting phenomena when light-matter interactions are considered, since the electric field forming such structures is capable of directly mediating many different light-matter interaction processes \cite{goldman2014light, canaguier2013force, de2017quantized, liu2019high}. 

A monochromatic electromagnetic wave propagating in free space along the $z$ direction can be described by an electric field distribution $\mathbf{E}(\mathbf{r})e^{i (k z-\omega t)}$. Here $\mathbf{E}$ is a complex 3-vector representing the envelope function of the field, $\mathbf{r}=(x,y,z)$ is the spatial coordinate, $t$ is the time, $\omega = c k$ is the angular frequency of the wave, with $c$ the speed of light in vacuum and $k$ the wavevector component along the $z$-direction. From the field distribution, one can define the electric-field-induced photonic spin density \cite{nye1987wave, barnett2010rotation, bliokh2015transverse, neugebauer2015measuring, bliokh2019geometric}:
\begin{equation}
    \mathbf{S} = \frac{\epsilon_0}{2\omega}\mathrm{Im}(\mathbf{E}^*\times\mathbf{E})
\end{equation}
Here $\epsilon_0$ is the vacuum permittivity. This is a vector perpendicular to the polarization ellipse of the electric field.

\begin{figure}
\includegraphics[width=\textwidth]{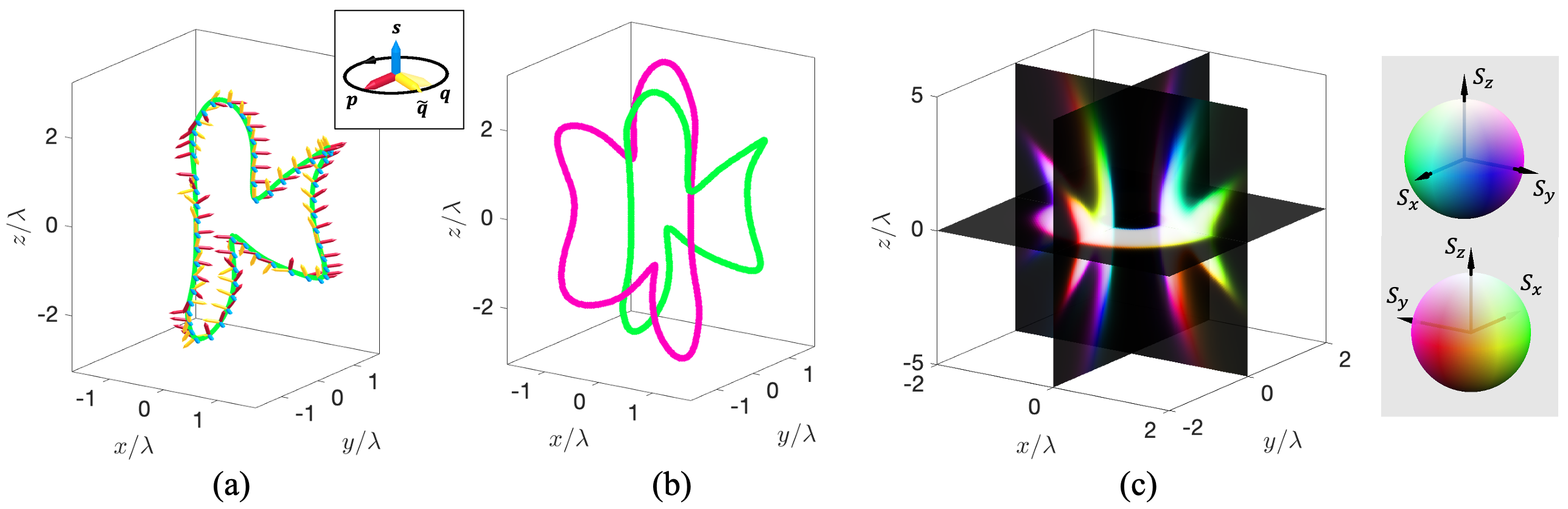}%
\caption{\label{fig_texture} Photonic Shankar skyrmion in monochromatic light. (a) The loop on which the spin densities are oriented along the $(1,-1,0)$ direction (green) and the texture of the triad along the loop. The triad rotates by $4\pi$ along the loop. The inset shows the definition of the triad. (b) Two loops on which the spin densities are oriented along the $(1,-1,0)$ direction (green) and the $(-1,1,0)$ direction (purple). (c) Distribution of the photonic spin density, which forms a hopfion texture. The color corresponds to the orientation of the spin density vector. Its correspondence is shown as the inset on the right side.}
\end{figure}

Using the electric field and the associated spin density allows one to define a triad at spatial locations where the field is elliptically polarized (inset of Fig.~\ref{fig_texture}a). We define quadrature vectors $\mathbf{P}=\frac{\mathbf{E}+\mathbf{E}^*}{2}$ and $\mathbf{Q}=\frac{\mathbf{E}-\mathbf{E}^*}{2i}$. The normalized 3-vector $\mathbf{p}=\mathbf{P}/|\mathbf{P}|$ forms one arm of the triad. From quadrature vector $\mathbf{Q}$, one can define another 3-vector perpendicular to $\mathbf{P}$ in the plane of the polarization ellipse, $\mathbf{\tilde{Q}} = \mathbf{Q} - \frac{\mathbf{P}\cdot\mathbf{Q}}{|\mathbf{P}|^2}\mathbf{P}$. The normalized 3-vector $\mathbf{\tilde{q}}=\mathbf{\tilde{Q}}/|\mathbf{\tilde{Q}}|$ forms the second arm of the triad. The third arm of the triad is formed by the normalized spin density $\mathbf{s}=\mathbf{S}/|\mathbf{S}|$. By construction, these three arms are mutually orthogonal to each other. For a given orientation of such a triad in the 3D space, one can uniquely associate with it an element of the 3D rotation group SO(3). Therefore, an SO(3) order parameter distribution can be associated with monochromatic electromagnetic waves. Note that the above procedure fails at locations where the electric field is linearly polarized. As we will see below, this leads to the defects of the order parameter distribution.

The SO(3) order parameter distribution leads to the existence of a topological quasiparticle in 3D space, known as the Shankar skyrmion \cite{shankar1977applications}. Such distributions are classified by the homotopy group $\pi_3(SO(3))=\mathbb{Z}$ \cite{mermin1979topological, han2017skyrmions}. The structure of a Shankar skyrmion of unity topological charge can be understood as follows. If one arm from the triad is chosen, one gets a vector field distribution in 3D space. The vector field forms a topologically non-trivial texture in 3D space, known as a hopfion \cite{tai2019three, kent2021creation, zheng2023hopfion, sugic2021particle, wang2023photonic}. In a concrete example that will be detailed later (Eq.~(\ref{shankarEfield})), we choose the normalized spin density $\mathbf{s}$ from the triad, its distribution forms a hopfion as shown in Fig.~\ref{fig_texture}c. Choosing the other two arms of the triad also lead to hopfions \cite{suppmat}. Hopfions can be viewed as concrete examples of the Hopf fibration \cite{hopf1931abbildungen, urbantke2003hopf}, which divides 3-spheres into linked loops. We illustrate two such loops corresponding to two distinct spin orientations in Fig.~\ref{fig_texture}b. Such loops are linked according to the structure of a spin texture hopfion (Fig.~\ref{fig_texture}c). On any one of such loops, the triad performs a $4\pi$ rotation along the loop, as illustrated in Fig.~\ref{fig_texture}a. Different from the usual vortices that contain singularities, inside such a loop of $4\pi$ rotation of the triad, there is no singularity. This texture is known as the Anderson-Toulouse vortex \cite{anderson1977phase, nakahara2018geometry}. This phenomenon originates from the homotopy group $\pi_1(SO(3))=\mathbb{Z}_2$, which shows that two charge-one vortices can annihilate each other, and a texture having $4\pi$ winding is homotopically equivalent to a trivial texture. 

Now we demonstrate how Shankar skyrmions can arise in monochromatic electromagnetic waves. We choose the envelope function $\mathbf{E}$ as:
\begin{equation}
    \mathbf{E} =  1.5\mathrm{\mathbf{LG}}_{(1,0)}^R -\mathrm{\mathbf{LG}}_{(0,0)}^R + 0.6[1,-i,0]^{\mathrm{T}} + 0.7\mathrm{\mathbf{LG}}_{(0,-2)}^L
    \label{shankarEfield}
\end{equation}
Here $\mathrm{\mathbf{LG}}^{R/L}_{(p,m)}$ are envelope functions of the Laguerre-Gaussian modes, but here we include both transverse and longitudinal components \cite{suppmat}. The superscript $R$($L$) denotes right (left) circular polarization, abbreviated as RCP (LCP). $p$ is the radial mode index, and $m$ is the azimuthal mode index. The Rayleigh ranges of all Laguerre-Gaussian modes are chosen to be $z_R=2\lambda$, where $\lambda$ is the wavelength of light. This corresponds to an extremely tightly focused beam. The third term in Eq. (\ref{shankarEfield}) represents a plane wave component. In practice, the plane wave can be replaced with a fundamental Gaussian beam with a large Rayleigh range.

\begin{figure}
\includegraphics[width=0.6\textwidth]{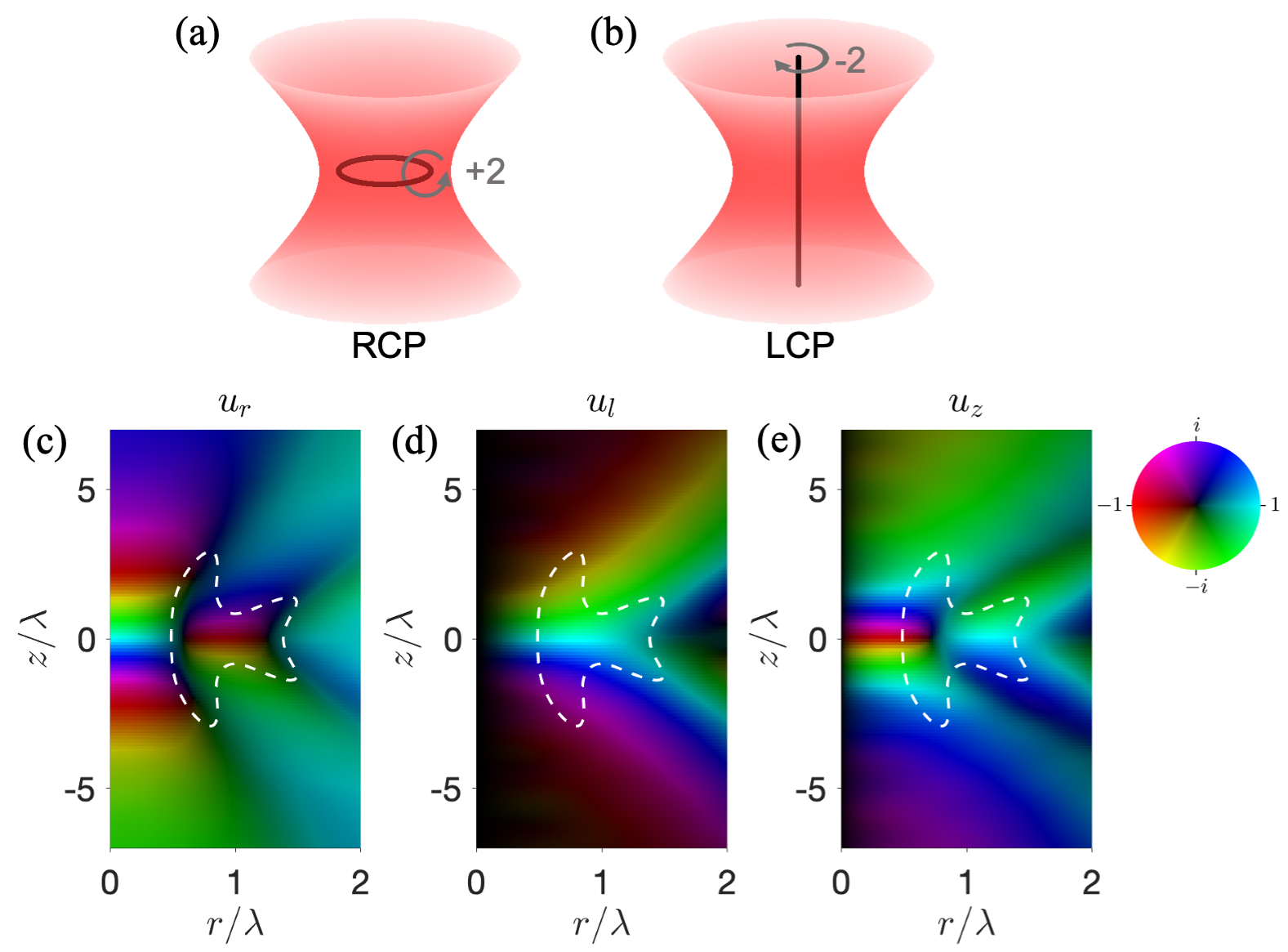}%
\caption{\label{fig_field} The field profiles of a Shankar skyrmion. (a) Schematic of the RCP envelope $u_r$, which contains a vortex ring of charge 2. (b) Schematic of the LCP envelope $u_l$, which contains a vortex line of charge 2. In (a) and (b) the arrows denote the direction of increasing phase. (c-e) The complex amplitude of $u_r$, $u_l$, and $u_z$ of the field defined in Eq.~(\ref{shankarEfield}). In each plot, the maximum amplitude is normalized to 1. The white dashed line denotes where $|u_r|=|u_l|$. The inset on the right shows the coloring scheme.}
\end{figure}

Such choices of beam profile originate from the geometric configurations of a Shankar skyrmion. Along the $z$ axis, the spin density points to the $-z$ direction and the triad performs a $4\pi$ rotation about the $z$ axis. Such a rotation of the triad arises from the Gouy phase of the RCP Laguerre-Gaussian modes, along with the RCP plane wave component, which ensures the total phase change of the field envelope along the $z$ axis is an integer multiple of $2\pi$. This $4\pi$ rotation of the triad is consistent with the formation of the Anderson-Toulouse vortex. This configuration can be schematically illustrated in Fig.~\ref{fig_field}a, where the $4\pi$ phase change is represented by a charge 2 vortex ring around the beam waist.

If we consider another loop where the spin density points exactly upwards, such a loop naturally circles the $z$ axis due to the structure of the spin hopfion. The triad again has to rotate $4\pi$ along such a loop. Such rotation can arise if the LCP mode possesses a vortex along the propagation direction with charge 2. This configuration is schematically illustrated in Fig.~\ref{fig_field}b.

One can also decompose the field envelope $\mathbf{E} = u_r(\hat{\mathbf{x}}-i\hat{\mathbf{y}})/\sqrt{2} + u_l(\hat{\mathbf{x}}+i\hat{\mathbf{y}})/\sqrt{2} + u_z\hat{\mathbf{z}}$, where $\hat{\mathbf{x}}$, $\hat{\mathbf{y}}$, and $\hat{\mathbf{z}}$ are the unit vectors of the coordinate frame. $u_r$, $u_l$, and $u_z$ are respectively the envelope functions of the RCP, the LCP, and the $z$ components. We plot the three envelope functions in Figs.~\ref{fig_field}c, \ref{fig_field}d, and \ref{fig_field}e. In Fig.~\ref{fig_field}c we see three vortices of charge 1 and one vortex of charge -1. This gives rise to a net vortex of charge 2 and is consistent with the schematic of Fig.~\ref{fig_field}a. The white dashed lines represent a loop where $|u_r| = |u_l|$. In the current example, we have chosen $z_R=2\lambda$, and such a loop encloses all the vortices and has a phase evolution of $4\pi$ along the loop, resulting in the creation of the Shankar skyrmion, in which the spin density does not vanish throughout the 3D space.

The choice of the $z_R$ above is an important consideration in the creation of the Shankar skyrmion in our configuration. If we choose a larger $z_R$, the loop encircles individual vortices separately, which leads to a phase change of $2\pi$ along each of the loops, which then gives rise to spin singularities. To see how spin singularities arise, we write the transverse field on the loop as:
\begin{equation}
\begin{aligned}
    \mathbf{E}_T &= (\mathbf{\hat{x}}-i\mathbf{\hat{y}})e^{i\phi(h)} + (\mathbf{\hat{x}}+i\mathbf{\hat{y}}) \\
    &= (2\cos{\frac{\phi(h)}{2}}\mathbf{\hat{e}}_r - 2\sin{\frac{\phi(h)}{2}}\mathbf{\hat{e}}_\theta)e^{i\frac{\phi(h)}{2}}
\end{aligned}
\end{equation}
Here $r=\sqrt{x^2+y^2}$ and $\theta=\arctan(y/x)$ are the cylindrical coordinates, $\mathbf{\hat{e}}_r$ and $\mathbf{\hat{e}}_\theta$ are unit vectors in the corresponding directions. $h$ is the coordinate along the loop. $\phi(h)$ is the phase difference between the RCP and LCP components. We see that when $\phi(h)$ increases by odd multiples of $2\pi$ along a given loop, regardless of the phase distribution of $E_z$, there exists at least one point on the loop where the phase of $\mathbf{E}_T$ and $E_z$ are the same or differ by $\pi$, and therefore the spin density vanishes. When $\phi(h)$ changes by $4n\pi$ along the loop, where $n$ is an integer, a necessary condition to avoid singularity on the loop is to have the phase change of $E_z$ along the loop to be $2n\pi$. Our example corresponds to the case where $n=1$, illustrated in Fig.~\ref{fig_field}e.

To calculate the topological charge of the Shankar skyrmion, we note that the triad, taking value in the space of SO(3), can be parametrized using a unit 4-vector $\mathfrak{q}=(w,\mathbf{v})$. Here $\mathbf{v}$ represents the rotation axis of the SO(3) element and $w$ parametrizes the rotation angle of the SO(3) element. The sign of the 4-vector can be chosen based on continuity. The topological charge is thus defined as \cite{zarzuela2019hydrodynamics}:
\begin{equation}
    Q = \frac{1}{2\pi^2}\int\mathrm{det}[\mathfrak{q}, \partial_x\mathfrak{q}, \partial_y\mathfrak{q}, \partial_z\mathfrak{q}]\mathrm{d}\mathbf{r}
\end{equation}
For the electric field distribution demonstrated in Eq. (\ref{shankarEfield}), $Q=-1$. Shankar skyrmion of the opposite charge can be created by switching the mode profile for left and right circular polarizations and changing the sign of the azimuthal index $m$ in the Laguerre-Gaussian modes.

\begin{figure}
\includegraphics[width=0.6\textwidth]{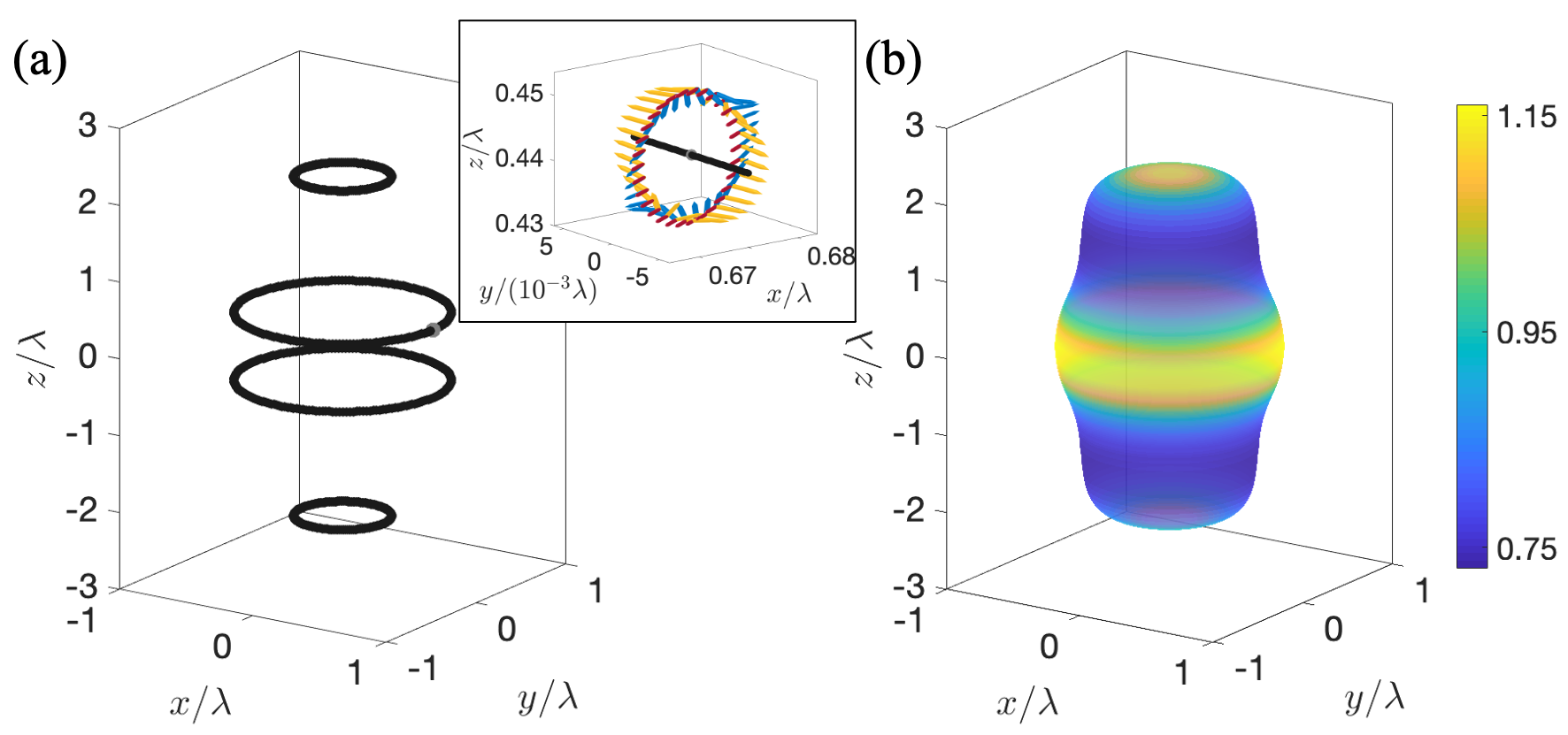}%
\caption{\label{fig_singularity} Photonic spin singularities during topological phase transition. (a) Singular rings in 3D space when $d=1.0$. The inset shows the texture of the triad around the singular line, near $(x,y,z)=(1,0,0)\lambda$. (b)  Singular surface in 4D space spanned by $xyz$ and $d$. The value of $d$ is represented by the color.}
\end{figure}

Now we proceed to demonstrate a few novel phenomena associated with such topological texture. We first discuss the topological phase transition between a Shankar skyrmion and the trivial texture. Such a transition can be induced by changing the coefficients in the superposition of modes in Eq. (\ref{shankarEfield}). We denote $d$ being the coefficient of the RCP plane wave component, initially taking the value of $0.6$. When we increase $d$ to be in the range of $[0.72, 1.16]$, we see singularity lines appear in space. In Fig.~\ref{fig_singularity}a, we have chosen $d=1.0$ and plotted the location of the singularity in 3D space. Such singularities correspond to locations where the electric field is linearly polarized, and hence the spin density vanishes, which does not allow the definition of a triad. Around the singularity, the triad forms a vortex texture, where $\mathbf{p}$ is nearly constant, and the vectors $\tilde{\mathbf{q}}$ and $\mathbf{s}$ form vortices or anti-vortices. In the inset of Fig.~\ref{fig_singularity}a, we show a case when anti-vortices are formed. Such defects are previously known as the $L^T$ lines or the spin singularities \cite{nye1987wave, freund2010optical, berry2001polarization, fang2025topological, shen2025free}, and are classified by the $\pi_1(SO(3))=\mathbb{Z}_2$ homotopy group.

When $d>1.16$, all the defect lines annihilate and we are left with a smooth texture of the triad but it is topologically trivial. Therefore, such defects form a 2D surface in the 4D space spanned by the $xyz$ coordinates and the parameter $d$. We call this surface the $L^T$ surface, illustrated in Fig.~\ref{fig_singularity}b, where the value of $d$ is represented by the color. Although the $L^T$ lines have been previously studied, their ability to form a 2D surface was not previously discussed. For a phase diagram when other parameters are varied, see supplementary material \cite{suppmat}.

The topological quasiparticle created here can also acquire propagation dynamics. It has been shown that one can transform a monochromatic beam into a non-diffracting wavepacket by adding spatiotemporal correlation \cite{kondakci2017diffraction, yessenov2023relativistic, wang2025spatiotemporal}. Specifically, we consider the following spatiotemporal correlation:
\begin{equation}
    v_g\cdot(k_z-k_0) = \omega' - ck_0
    \label{STcorr}
\end{equation}
$\mathbf{k}=(k_x, k_y, k_z)$ and $\omega'$ is the wavevector and the angular frequency of the plane wave component. They satisfy $\omega'=c\sqrt{k_x^2+k_y^2+k_z^2}$. $v_g$ is an arbitrarily chosen group velocity, and $k_0$ is a fixed constant chosen equal to $k$ in the following.

\begin{figure}
\includegraphics[width=0.6\textwidth]{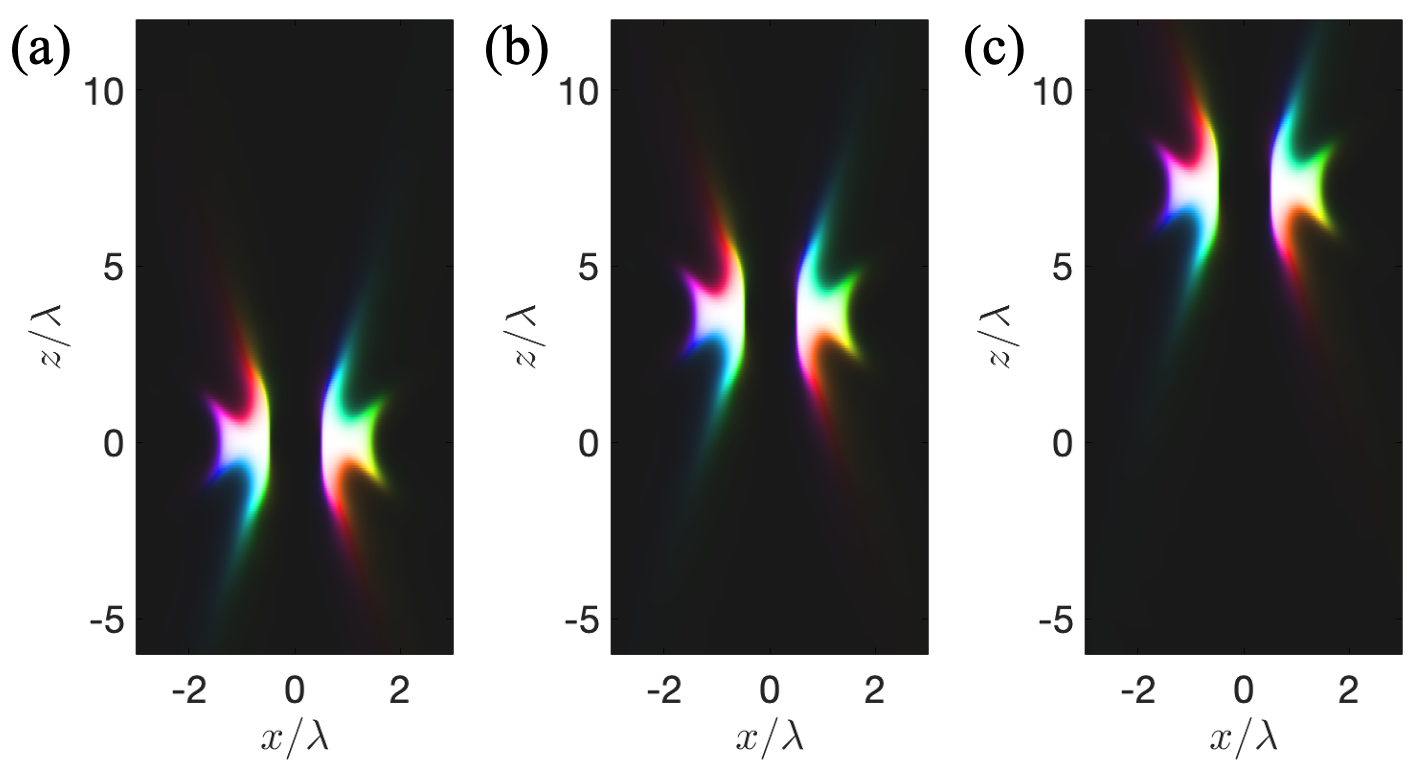}%
\caption{\label{fig_bullet} Flying topological quasiparticle. The photonic spin density on the $xz$-plane for a photonic Shankar skyrmion moving at velocity $v_g=0.3c$ along the $z$-axis is shown at time $t=(0, 12, 24)\lambda/c$, in (a), (b), and (c) respectively. The spin orientations are represented by various colors using the same scheme as Fig.~\ref{fig_texture}.}
\end{figure}

The flying Shankar skyrmion can be constructed as follows. Starting from the field distribution in Eq. (\ref{shankarEfield}), one can decompose such a field by its Fourier components:
\begin{equation}
    \mathbf{E}e^{ikz} = \int g(k_x,k_y)e^{i(k_x x + k_y y + k_z z)}\hat{\mathbf{e}}(k_x,k_y)dk_xdk_y
\end{equation}
Here $g(k_x,k_y)$ is the amplitude of each plane wave component, and $\hat{\mathbf{e}}$ is the polarization basis for each plane wave. We replace the propagation phase $e^{ik_z z}$ in the integral where $k_z = \sqrt{\omega^2/c^2-k_x^2-k_y^2}$ with a phase factor $e^{i k_z z-i\omega' t}$ where the wavevector and angular frequency satisfies the spatiotemporal correlation described by Eq. (\ref{STcorr}). The resulting field distribution then describes a Shankar skyrmion wave packet in the 3D real space, propagating along the $z$ axis at velocity $v_g$.

The order parameter in such polychromatic waves can be defined from its real-valued electric field $\mathbf{E}_r$ and its real-valued vector potential $\mathbf{A}_r$. Together with the spin density in such waves, defined as $\mathbf{S}=\epsilon_0 \mathbf{E}_r\times \mathbf{A}_r$ \cite{li2009spin, barnett2010rotation, bialynicki2011canonical, mansuripur2011spin, gutierrez2021optical}, the SO(3) order parameter can be defined \cite{suppmat}. These definitions are generalizations of the monochromatic case described above. In Fig.~\ref{fig_bullet}, we plot the distribution of the spin density for such waves at time $t=(0, 12, 24)\lambda/c$. We see that the spin distribution, and therefore the Shankar skyrmion, propagates in the $z$ direction without diffraction at a group velocity of $0.3c$. We note that such spatiotemporal correlation can be experimentally implemented by established pulse shaping methods \cite{yessenov2024experimental} or by spatiotemporal refraction \cite{wang2025spatiotemporal}. The group velocity of such a topological quasiparticle can be further controlled by varying the refractive index along its propagation direction \cite{bhaduri2020anomalous, wang2025spatiotemporal}.

The topological texture we introduced above can be viewed as a structured elliptical polarization, solely determined by the electric field of the wave. Given that the electric field and electric dipole moments usually provide the dominant contribution in many light-matter interaction processes, such topological texture may directly lead to many new opportunities, such as in manipulating the internal degrees of freedom of various media or microscopic objects. In ultracold gases, it was shown that a monochromatic elliptical light field detuned from atomic transitions imposes an effective magnetic field on the atoms, given by $\mathbf{B}_{\mathrm{eff}}=i\alpha\mathbf{E^*}\times\mathbf{E}$
where $\alpha$ is a real constant determined by the properties of the atomic transitions \cite{goldman2014light, zhu2013absolute}. The photonic Shankar skyrmion may directly impose an effective magnetic field, in the form of a hopfion, onto the ultracold gases. Apart from the hopfion texture, one can also show that $\mathbf{B}_{\mathrm{eff}}$ is not divergence-free. Such effective magnetic fields are therefore different from real magnetic fields in free space which must be divergence free, and may lead to new opportunities in quantum emulation \cite{pietila2009creation, ray2014observation}. In the strong-field regime, such structured elliptical polarization may also lead to the creation of a structured magnetic field in plasma via the inverse-Faraday effect \cite{liu2019high}.

In condensed matter physics, electronic states can be modified by photonic spin \cite{mak2012control, mciver2020light}, therefore such light may potentially create structured excitations of electrons. The topology of light may also be transferred to electronic degrees of freedom \cite{olbrich2009observation, de2017quantized}, leading to hybrid topological quasiparticles of light and matter. Photonic spin density also applies torque to nanoparticles  \cite{garces2003observation, canaguier2013force}, thus such field configuration may also find application in optical manipulation.

In summary, we have demonstrated a new electromagnetic topological quasiparticle, the Shankar skyrmion of light. We discussed how a static topological quasiparticle can be constructed in monochromatic fields, and a propagating topological quasiparticle can be constructed by transforming the monochromatic field through adding spatiotemporal correlation. The transition between the topologically nontrivial configuration and the trivial configuration leads to the appearance of spin singularities, which form a 2D surface in 4D space. The topological texture arises solely from the electric field distribution of the electromagnetic wave. Such topological quasiparticles may lead to new phenomena in manipulating electronic states in solids or charged particles in plasma, serve as a tool for quantum emulation, or be used in optical manipulation.

\section*{Acknowledgements}
This work is supported by the Office of Naval Research (Grant No. N00014-20-1-2450) and by the Army Research Office (Grant No. W911NF-24-2-0170).

\end{document}